\documentstyle{aipproc}

\newcommand{\beq}{\begin{equation}}
\newcommand{\eeq}{\end{equation}}
\def\bea{\begin{eqnarray}}
\def\eea{\end{eqnarray}}
\def\nn{\nonumber}
\def\sss{\scriptscriptstyle}

\def\bd{B_d^0}
\def\bdbar{{\overline{B_d^0}}}
\def\bs{B_s^0}
\def\bsbar{{\overline{B_s^0}}}
\def\barp{{\raise.35ex\hbox{${\sss (}$}}---{\raise.35ex\hbox{${\sss )}$}}}
\def\bdbarp{\hbox{$B_d$\kern-1.4em\raise1.4ex\hbox{\barp}}}
\def\bsbarp{\hbox{$B_s$\kern-1.4em\raise1.4ex\hbox{\barp}}}
\def\ks{K_{\sss S}}

\def\kbar{{\overline{K^0}}}

\def\roughly#1{\mathrel{\raise.3ex\hbox{$#1$\kern-.75em\lower1ex\hbox{$\sim$}}}}

\def\Abar{\bar A}
\def\Ptilde{{\tilde P}}
\def\epjc#1#2#3{{\it Eur.\ Phys.\ J.}\ {\bf C#1}, #3 (19#2)} 

\def\npb#1#2#3{{\it Nucl.\ Phys.}\ {\bf B#1}, #3 (19#2)}
\def\plb#1#2#3{{\it Phys.\ Lett.}\ {\bf #1B}, #3 (19#2)}
\def\prd#1#2#3{{\it Phys.\ Rev.}\ {\bf D#1}, #3 (19#2)}
\def\prl#1#2#3{{\it Phys.\ Rev.\ Lett.}\ {\bf #1}, #3 (19#2)}
\def\zpc#1#2#3{{\it Zeit.\ Phys.}\ {\bf C#1}, #3 (19#2)} 
\def\ijmp#1#2#3{{\it Int.\ J.\ Mod.\ Phys.}\ {\bf A#1}, #3 (19#2)}

\newread\epsffilein % file to \read
\newif\ifepsffileok % continue looking for the bounding box?
\newif\ifepsfbbfound % success?
\newif\ifepsfverbose % report what you're making?
\newdimen\epsfxsize % horizontal size after scaling
\newdimen\epsfysize % vertical size after scaling
\newdimen\epsftsize % horizontal size before scaling
\newdimen\epsfrsize % vertical size before scaling
\newdimen\epsftmp % register for arithmetic manipulation
\newdimen\pspoints % conversion factor
\def\epsfbox#1{\global\def\epsfllx{72}\global\def\epsflly{72}%
 \global\def\epsfurx{540}\global\def\epsfury{720}%
 \def\lbracket{[}\def\testit{#1}\ifx\testit\lbracket
 \let\next=\epsfgetlitbb\else\let\next=\epsfnormal\fi\next{#1}}%
\def\epsfgetlitbb#1#2 #3 #4 #5]#6{\epsfgrab #2 #3 #4 #5 .\\%
 \epsfsetgraph{#6}}%
\def\epsfnormal#1{\epsfgetbb{#1}\epsfsetgraph{#1}}%
\def\epsfgetbb#1{%
\openin\epsffilein=#1
\ifeof\epsffilein\errmessage{I couldn't open #1, will ignore it}\else
 {\epsffileoktrue \chardef\other=12
 \def\do##1{\catcode`##1=\other}\dospecials \catcode`\ =10
 \loop
 \read\epsffilein to \epsffileline
 \ifeof\epsffilein\epsffileokfalse\else
 \expandafter\epsfaux\epsffileline:. \\%
 \fi
 \ifepsffileok\repeat
 \ifepsfbbfound\else
 \ifepsfverbose\message{No bounding box comment in #1; using defaults}\fi\fi
 }\closein\epsffilein\fi}%
\def\epsfclipstring{}% do we clip or not? If so,
\def\epsfsetgraph#1{%
 \epsfrsize=\epsfury\pspoints
 \advance\epsfrsize by-\epsflly\pspoints
 \epsftsize=\epsfurx\pspoints
 \advance\epsftsize by-\epsfllx\pspoints
 \epsfxsize\epsfsize\epsftsize\epsfrsize
 \ifnum\epsfxsize=0 \ifnum\epsfysize=0
 \epsfxsize=\epsftsize \epsfysize=\epsfrsize
 \epsfrsize=0pt
 \else\epsftmp=\epsftsize \divide\epsftmp\epsfrsize
 \epsfxsize=\epsfysize \multiply\epsfxsize\epsftmp
 \multiply\epsftmp\epsfrsize \advance\epsftsize-\epsftmp
 \epsftmp=\epsfysize
 \loop \advance\epsftsize\epsftsize \divide\epsftmp 2
 \ifnum\epsftmp>0
 \ifnum\epsftsize<\epsfrsize\else
 \advance\epsftsize-\epsfrsize \advance\epsfxsize\epsftmp \fi
 \repeat
 \epsfrsize=0pt
 \fi
 \else \ifnum\epsfysize=0
 \epsftmp=\epsfrsize \divide\epsftmp\epsftsize
 \epsfysize=\epsfxsize \multiply\epsfysize\epsftmp
 \multiply\epsftmp\epsftsize \advance\epsfrsize-\epsftmp
 \epsftmp=\epsfxsize
 \loop \advance\epsfrsize\epsfrsize \divide\epsftmp 2
 \ifnum\epsftmp>0
 \ifnum\epsfrsize<\epsftsize\else
 \advance\epsfrsize-\epsftsize \advance\epsfysize\epsftmp \fi
 \repeat
 \epsfrsize=0pt
 \else
 \epsfrsize=\epsfysize
 \fi
 \fi
 \ifepsfverbose\message{#1: width=\the\epsfxsize, height=\the\epsfysize}\fi
 \epsftmp=10\epsfxsize \divide\epsftmp\pspoints
 \vbox to\epsfysize{\vfil\hbox to\epsfxsize{%
 \ifnum\epsfrsize=0\relax
 \includegraphics{#1}%
 \else
 \epsfrsize=10\epsfysize \divide\epsfrsize\pspoints
 \includegraphics{#1}%
 \fi
 \hfil}}%
\global\epsfxsize=0pt\global\epsfysize=0pt}%
 {\catcode`\%=12 \global\let\epsfpercent=%\global\def\epsfbblit{%BoundingBox}}%
\long\def\epsfaux#1#2:#3\\{\ifx#1\epsfpercent
 \def\testit{#2}\ifx\testit\epsfbblit
 \epsfgrab #3 . . . \\%
 \epsffileokfalse
 \global\epsfbbfoundtrue
 \fi\else\ifx#1\par\else\epsffileokfalse\fi\fi}%
\def\epsfempty{}%
\def\epsfgrab #1 #2 #3 #4 #5\\{%
\global\def\epsfllx{#1}\ifx\epsfllx\epsfempty
 \epsfgrab #2 #3 #4 #5 .\\\else
 \global\def\epsflly{#2}%
 \global\def\epsfurx{#3}\global\def\epsfury{#4}\fi}%
\def\epsfsize#1#2{\epsfxsize}
\begin{document}

\begin{flushright}
UdeM-GPP-TH-99-64
\end{flushright}

\title{Looking for New Physics in $\bd$-$\bdbar$ Mixing\thanks{Seminar
    given at {\it MRST '99: High Energy Physics at the Millenium},
    Carleton University, Ottawa, Canada, May 1999. Talk based on work
    done in collaboration with A. Ali, N.  Sinha and R. Sinha, and
    C.S. Kim and T. Yoshikawa.}}

\author{David London}

\address{Laboratoire de Ren\'e J.-A. L\'evesque, Universit\'e de
Montr\'eal\\
C.P. 6128, succ.\ centre-ville, Montr\'eal, QC, Canada}

\maketitle

\begin{abstract}
  There are variety of methods which directly test for the presence of
  new physics in the $b\to s$ flavour-changing neutral current (FCNC),
  but none which cleanly probe new physics in the $b\to d$ FCNC. One
  possible idea is to compare the weak phase of the $t$-quark
  contribution to the $b\to d$ penguin, which is $-\beta$ in the SM,
  with that of $\bd$-$\bdbar$ mixing ($-2\beta$ in the SM). In this
  talk I show that, in fact, it is impossible to measure the weak
  phase of the $t$-quark penguin, or indeed any penguin contribution,
  without theoretical input. However, if one makes a single assumption
  involving the hadronic parameters, it is possible to obtain the weak
  phase. I discuss how one can apply such an assumption to the
  time-dependent decays $\bd(t) \to K^0\kbar$ and $\bs(t) \to\phi\ks$
  in order to detect new physics in the $b\to d$ FCNC.
\end{abstract}

In the coming years, experiments at B-factories, HERA-B and hadron
colliders will measure CP-violating asymmetries in $B$ decays
\cite{BCPasym}. As always, the goal is to test the predictions of the
standard model (SM). If we are lucky, there will be an inconsistency
with the SM, thereby revealing the presence of new physics.

In the SM, CP violation is due to a complex phase in the
Cabibbo-Kobayashi-Maskawa (CKM) mixing matrix. In the Wolfenstein
parametrization of the CKM matrix \cite{Wolfenstein}, only the
elements $V_{ub}$ and $V_{td}$ have non-negligible phases:
\beq
V_{\sss CKM} = \left( \matrix{ 
1 - {1\over 2} \lambda^2 & \lambda & A \lambda^3 (\rho - i \eta) \cr
- \lambda & 1 - {1\over 2} \lambda^2 & A \lambda^2 \cr
A \lambda^3 ( 1 - \rho - i \eta) & - A \lambda^2 & 1 \cr} \right).
\eeq
It is convenient to parametrize $V_{ub}$ and $V_{td}$ as follows:
\beq
V_{ub} = | V_{ub} | e^{-i \gamma} ~~,~~~~~
V_{td} = | V_{td} | e^{-i \beta} ~~.
\eeq
Even though these elements are written in terms of two complex phases
$\beta$ and $\gamma$, it must be remembered that in fact there is only a
single phase $\eta$ in the CKM matrix; if $\eta$ were to vanish, both
$\beta$ and $\gamma$ would vanish as well.

The phase information in the CKM matrix can be elegantly displayed
using the unitarity triangle \cite{PDG}. The orthogonality of the
first and third columns gives
\beq
V_{ud} V_{ub}^* + V_{cd} V_{cb}^* + V_{td} V_{tb}^* = 0 ~,
\label{unitarityrelation}
\eeq
which is a triangle relation in the complex $\rho$-$\eta$ plane, shown in
Fig.~1. The angles $\beta$ and $\gamma$ are two of the interior angles of
the unitarity triangle, with the third angle $\alpha$ satisfying $\alpha +
\beta + \gamma = \pi$.

% This is Figure 1
\begin{figure}
\vskip -1.0truein
\centerline{\epsfxsize 3.5 truein \epsfbox {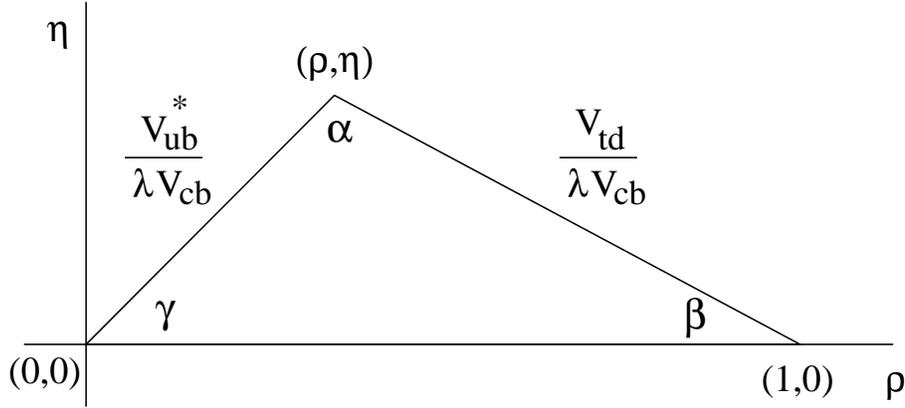}}
\vskip -1.2truein
\caption{The unitarity triangle. The angles $\alpha$, $\beta$ and $\gamma$
can be measured via CP violation in the $B$ system.}
\label{triangle}
\end{figure}

There are a variety of constraints on the unitarity triangle coming
from (i) the extraction of $|V_{cb}|$ and $|V_{ub}|$ from semileptonic
$B$ decays, (ii) the measurements of $|V_{td}|$ and $|V_{ts}|$ in
$\bd$-$\bdbar$ and $\bs$-$\bsbar$ mixing, and (iii) CP violation in
the kaon system ($\epsilon$). Unfortunately, there are substantial
theoretical uncertainties in all of these constraints. For example,
the theoretical expressions for $\epsilon$ and $\bd$-$\bdbar$ mixing
depend respectively on the bag parameter $B_{\sss K} = 0.94 \pm 0.15$
and $f_{B_d} \sqrt{B_{B_d}} = 215 \pm 40$ MeV. The estimates of the
magnitudes of these errors, which lie in the range 15--20\%, come
mainly from lattice calculations. Combining the experimental errors
and theoretical uncertainties in quadrature \cite{AliLon}, the
presently-allowed region of the unitarity triangle is shown in Fig.~2.
Due to the theoretical uncertainties, we do not have precise SM
predictions for the CP phases $\alpha$, $\beta$ and $\gamma$: instead,
these phases can take a range of values.

% This is Figure 2
\begin{figure}
\vskip -1.2truein
\centerline{\epsfxsize 4.0 truein \epsfbox {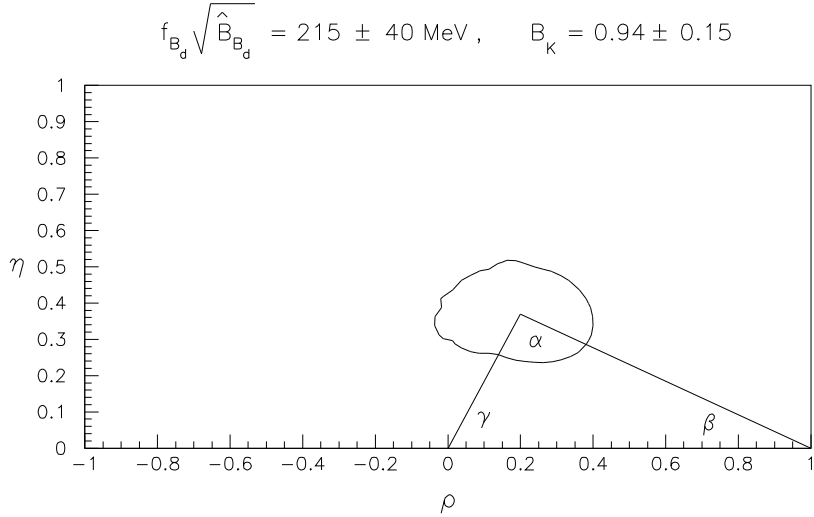}}
\vskip -1.4truein
\caption{Allowed region (95\% C.L.) in the $\rho$-$\eta$ plane, from a
simultaneous fit to all experimental and theoretical data. The theoretical
errors are treated as Gaussian for this fit. The triangle shows the best
fit.} 
\label{totalfit}
\end{figure}

Since the hope is to find physics beyond the SM, the first question to
be answered is: how can new physics affect the CP-violating
asymmetries? There are two possible ways: the new physics can affect
$B$ decays or $B$ mixing. Now, most $B$ decays are dominated by a
$W$-mediated tree-level diagram. In most models of new physics, there
are no contributions to $B$ decays which can compete with the SM.
Thus, in general, the new physics cannot significantly affect the
decays\footnote{There is an exception: if the decay process is
  dominated by a penguin diagram, rather than a tree-level diagram,
  then new physics {\it can} significantly affect the decay, see
  Refs.~\cite{BCPasym,NPpenguins}.}. However, the CP asymmetries {\it
  can} be affected if there are new contributions to
$B^0$-${\overline{B^0}}$ mixing \cite{NPBmixing}. The presence of such
new-physics contributions will affect the extraction of $V_{td}$ and
$V_{ts}$. And if there are new phases, the measurements of $\alpha$,
$\beta$ and $\gamma$ will also be affected. Thus, new physics enters
principally through new contributions to $B^0$-${\overline{B^0}}$
mixing \cite{Bnewphysics}.

Unfortunately, this creates a bit of a problem. $B$-factories such as
BaBar and Belle will measure $\alpha$, $\beta$ and $\gamma$ via
$\bd(t) \to \pi^+ \pi^-$ (or $\rho\pi$ \cite{Dalitz}), $\bd(t) \to
\Psi\ks$, and $B^\pm \to D K^\pm$ \cite{BtoDK}, respectively. Note
that only the first two decays involve $B^0$-${\overline{B^0}}$
mixing. Thus, if there is new physics, only the measurements of
$\alpha$ and $\beta$ will be affected. However, they will be affected
in opposite directions \cite{NirSilv}. That is, in the presence of a
new-physics phase $\phi_{\sss NP}$, the CP angles are changed as
follows: $\alpha \to \alpha + \phi_{\sss NP}$ and $\beta \to \beta -
\phi_{\sss NP}$.  The key point is that $\phi_{\sss NP}$ cancels in
the sum $\alpha + \beta + \gamma$, so that this sum is {\it
  insensitive} to the new physics, i.e.\ $B$-factories will always
find $\alpha + \beta + \gamma = \pi$.  (Note that hadron colliders do
not suffer from the same problem -- if $\gamma$ is measured in $\bs(t)
\to D_s^\pm K^\mp$ \cite{BstoDsK}, then $\alpha + \beta + \gamma \ne
\pi$ can be found if there is new physics in $\bs$-$\bsbar$ mixing.)

Thus, $B$-factories cannot discover new physics via $\alpha + \beta +
\gamma \ne \pi$. Still, new physics can be found if the measurements
of the angles are inconsistent with the measurements of the sides.
However:
\begin{enumerate}
  
\item the allowed region of the unitarity triangle is still fairly
  large. It is conceivable that even in the presence of new physics,
  the triangle as constructed from the angles $\alpha$, $\beta$ and
  $\gamma$ will still lie within the allowed region;
  
\item even if the $\alpha$-$\beta$-$\gamma$ triangle lies outside the
  allowed region, is this evidence of new physics, or have we
  underestimated the theoretical uncertainties which go into the
  constraints of the unitarity triangle (Fig.~\ref{totalfit})?

\end{enumerate}
The point is: ideally, we would like cleaner, more direct tests of the
SM in order to probe for the presence of new physics.

In fact, there are such direct tests:
\begin{enumerate}
  
\item $B^\pm \to D K^\pm$ vs.\ $\bs(t) \to D_s^\pm K^\mp$: in the SM,
  both of these CP asymmetries measure $\gamma$. If there is a
  discrepancy between the value of $\gamma$ as extracted from these
  two decays, this points to new physics in $\bs$-$\bsbar$ mixing.
  
\item $\bd(t) \to \Psi\ks$ vs.\ $\bd(t) \to \phi\ks$: in the SM, both
  of these decays measure $\beta$. A discrepancy implies new physics
  in the $b\to s$ penguin \cite{NPpenguins}.
  
\item $\bs(t)\to\Psi\phi$: in the SM, the CP asymmetry in this decay
  vanishes (to a good approximation). If this CP asymmetry is found to
  be nonzero, this again indicates the presence of new physics in
  $\bs$-$\bsbar$ mixing.

\end{enumerate}
There are thus several direct tests for new physics. However, note:
all of these tests probe new physics in either $\bs$-$\bsbar$ mixing
or the $b\to s$ penguin, i.e.\ in the $b\to s$ flavour-changing
neutral current (FCNC).

So this raises the question: are there any direct tests of new physics
in the $b\to d$ FCNC?

Consider pure $b\to d$ penguin decays such as $\bd \to K^0\kbar$ or
$\bs\to\phi\ks$. Such decays involve up-type quarks in the loop. If
the $t$-quark contribution dominated, then the $b\to d$ penguin
amplitude would be proportional to $V_{tb}^* V_{td}$. Recalling that
the weak phase of $\bd$-$\bdbar$ mixing is $-2\beta$ and that the weak
phase of $V_{td}$ is $-\beta$, in such a case the SM would predict
that (i) the CP asymmetry in $\bd(t) \to K^0\kbar$ vanishes, and (ii)
the CP asymmetry in $\bs(t) \to\phi\ks$ measures $\sin 2\beta$
\cite{penguins}. Any discrepancy between measurements of these CP
asymmetries and their predictions would thus imply that there is new
physics in either $\bd$-$\bdbar$ mixing or the $b\to d$ penguin, i.e.\ 
in the $b \to d$ FCNC. (In the second decay, new physics in
$\bs$-$\bsbar$ mixing could also come into play, but that can be
established independently, as discussed above).

However, $b\to d$ penguins are {\it not} dominated by the internal
$t$-quark. The contributions of the $u$- and $c$-quarks can be as
large as 20--50\% of that of the $t$-quark \cite{ucquark}. In this
case, the above predictions of the SM no longer hold, so that one
cannot test for new physics in the $b\to d$ FCNC in this way.

So this raises a new question: are there ways of cleanly measuring the
weak phase of the $t$-quark contribution to the $b\to d$ penguin?
Unfortunately, the answer to this question is {\it no} \cite{LSS}.

To see this, consider the general form of the amplitude for the $b\to
d$ penguin. There are three terms, corresponding to the contributions
of the three internal up-type quarks:
\beq
P = P_u \, V_{ub}^* V_{ud} + P_c \, V_{cb}^* V_{cd} + P_t \, 
V_{tb}^* V_{td} ~,
\label{bdpenguin}
\eeq
and recall that $V_{ub} \sim e^{-i \gamma}$ and $V_{td} \sim e^{-i
  \beta}$.

Using the unitarity relation of Eq.~\ref{unitarityrelation}, the
$u$-quark piece can be eliminated in the above equation, allowing us
to write
\beq
P = {\cal P}_{cu} \, e^{i\delta_{cu}} + {\cal P}_{tu} \, e^{i\delta_{tu}} 
e^{-i \beta} ~,
\label{peng1}
\eeq
where $\delta_{cu}$ and $\delta_{tu}$ are strong phases. Now imagine
that there were a method in which a series of measurements allowed us
to cleanly extract $\beta$ using the above expression. In this case,
we would be able to express $-\beta$ as a function of the observables.

On the other hand, we can instead use the unitarity relation to
eliminate the $t$-quark contribution in Eq.~\ref{bdpenguin}, yielding
\beq
P = {\cal P}_{ct} \, e^{i\delta_{ct}} + {\cal P}_{ut} \, e^{i\delta_{ut}} 
e^{i \gamma} ~.
\label{peng2}
\eeq
Comparing Eqs.~\ref{peng1} and \ref{peng2}, we see that they have the
same form. Thus, the same method which allowed us to extract $-\beta$
from Eq.~\ref{peng1} should be applicable to Eq.~\ref{peng2}, allowing
us to obtain $\gamma$. That is, we would be able to write $\gamma$ as
{\it the same function} of the observables as was used for $-\beta$
above! But this implies that $-\beta = \gamma$, which clearly doesn't
hold in general.

Due to the ambiguity in the parametrization of the $b\to d$ penguin
--- which I will refer to as the {\it CKM ambiguity} --- we therefore
conclude that one cannot cleanly extract the weak phase of any penguin
contribution. Indeed, it is {\it impossible} to cleanly test for the
presence of new physics in the $b\to d$ FCNC.

Nevertheless, it is interesting to examine some candidate methods and
see how they fail. For example, consider the time-dependent rate for
the decay $\bd(t) \to K^0\kbar$. This can be written
\bea
\Gamma(\bd(t) \to K^0\kbar) & = & e^{-\Gamma t} 
\left[ {|A|^2 + |\Abar|^2 \over 2} 
+ {|A|^2 - |\Abar|^2 \over 2} \cos (\Delta M t) \right. \nn\\
& & \qquad\qquad\qquad\qquad 
- {\rm Im} \left( e^{-2i\beta} {A}^* \Abar\right) \sin (\Delta M t) 
\biggr],
\label{timedeprate}
\eea
where $A \equiv A(\bd \to K^0\kbar)$ and $\Abar \equiv A(\bdbar \to
K^0\kbar)$. The measurement of this time-dependent decay rate allows
one to extract the magnitudes and relative phase of $e^{i\beta} A$ and
$e^{-i\beta} \Abar$. Using the form of the $b\to d$ penguin given in
Eq.~\ref{peng1}, we have
\beq
  e^{i\beta} A = e^{i\beta} \left[ {\cal P}_{cu} \, e^{i\delta_{cu}} 
+ {\cal P}_{tu} \, e^{i\delta_{tu}} e^{-i \beta'} \right],
\eeq
where in the last term I have written the weak phase as $\beta'$ to
allow for the possibility of new physics in the $b\to d$ FCNC. There
are thus 5 measurable parameters: ${\cal P}_{cu}$, ${\cal P}_{tu}$,
$\delta_{cu}-\delta_{tu}$, $\beta$, and $\theta_{\sss NP} \equiv
\beta'-\beta$. However, there are only 4 measurements: the
coefficients of the 3 time-dependent functions [1, $\cos(\Delta Mt)$,
$\sin(\Delta Mt)$] in Eq.~\ref{timedeprate}, and one independent
measurement of $\beta$. Therefore, as argued above, there are not
enough measurements to determine all the theoretical parameters. More
to the point, there is one more theoretical unknown than there are
measurements.

In fact, one can examine a variety of other techniques: $B\to \pi\pi$
isospin analysis \cite{isospin}, Dalitz plot analysis of $B \to 3\pi$
\cite{Dalitz}, angular analysis of the decay of a neutral $B$-meson to
two vector mesons \cite{helicity}, and a combined isospin $+$ angular
analysis of $B\to \rho\rho$. In all cases there is one more unknown
than there are measurements. From this we can therefore conclude the
following: due to the CKM ambiguity, if we wish to test for the
presence of new physics in the $b\to d$ FCNC by comparing the weak
phase of $\bd$-$\bdbar$ mixing with that of the $t$-quark contribution
to the $b\to d$ penguin, it is necessary to make a single assumption
about the theoretical (hadronic) parameters describing the decay.

As an example of such an assumption, consider again the two decays
$\bd(t) \to K^0\kbar$ and $\bs(t) \to\phi\ks$. Recall that we can
write the $\bd \to K^0\kbar$ amplitude as
\beq 
e^{i\beta} A_d^{\sss K^0\kbar} = {\cal P}_{cu} \, e^{i\delta_{cu}}
e^{i\beta} + {\cal P}_{tu} \, e^{i\delta_{tu}} e^{-i (\beta' - \beta)} ~.
\eeq
Assuming that there is no new physics in $\bs$-$\bsbar$ mixing, we can
write the $\bs \to\phi\ks$ amplitude as
\beq
A_s^{\sss\phi\ks} = 
  {\tilde{\cal P}}_{cu} \, e^{i{\tilde\delta}_{cu}} 
+ {\tilde{\cal P}}_{tu} \, e^{i{\tilde\delta}_{tu}} e^{-i \beta'} ~.
\eeq
The tildes are added to distinguish the parameters in the decay $\bs
\to\phi\ks$ from those in $\bd \to K^0\kbar$. There are two reasons.
First, in the $\bs$ decay, we have a spectator $s$-quark instead of a
$d$-quark. And second, there are colour-allowed electroweak penguin
contributions to $\bs \to\phi\ks$ while there are none in $\bd \to
K^0\kbar$.

{}From the above, we see that there are 8 theoretical parameters
describing these two decays: ${\cal P}_{cu}$, ${\cal P}_{tu}$,
${\tilde{\cal P}}_{cu}$, ${\tilde{\cal P}}_{tu}$, $\beta$, $\beta'$,
$\delta_{cu}-\delta_{tu}$, and
${\tilde\delta}_{cu}-{\tilde\delta}_{tu}$. However there are only 7
experimental measurements: the magnitudes and relative phase of
$e^{i\beta} A_d^{\sss K^0\kbar}$ and $e^{-i\beta} \Abar_d^{\sss
  K^0\kbar}$, the magnitudes and relative phase of $A_s^{\sss\phi\ks}$ and
$\Abar_s^{\sss\phi\ks}$, and an independent measurement of $\beta$. If we
wish to determine the theoretical parameters, we therefore need to
make an assumption.

In Ref.~\cite{KLY2}, the following assumption is made. Defining $r
\equiv {\cal P}_{cu}/{\cal P}_{tu}$ and ${\tilde r} \equiv
{\tilde{\cal P}}_{cu} / {\tilde{\cal P}}_{tu}$, it is assumed that $r
= {\tilde r}$. How good is this assumption? Writing
\beq
r = \left\vert { P_c - P_u \over P_t - P_u + P_{\sss EW}^{\sss C}}
\right\vert ~,~~~ {\tilde r} = \left\vert { \Ptilde_c - \Ptilde_u \over
\Ptilde_t - \Ptilde_u + \Ptilde_{\sss EW} + \Ptilde_{\sss EW}^{\sss C} } 
\right\vert ~,
\eeq
we note the following. Since the spectator-quark effects cancel in the
ratio in ${\tilde r}$, the principle difference between $r$ and
${\tilde r}$ is due to the presence of the colour-allowed electroweak
penguin contribution in the denominator of ${\tilde r}$. Since
$\Ptilde_{\sss EW}/\Ptilde_t \simeq 20\%$, we therefore conclude that
$r$ and ${\tilde r}$ are equal to within roughly 20\%. Taking $r =
{\tilde r}$ is therefore a reasonable assumption.

With this assumption, we now have an equal number of theoretical
unknowns and experimental measurements, and can therefore solve for
$\beta$ and $\beta'$ independently. In this way we can test for the
presence of new physics in the $b\to d$ FCNC. Note also that the
assumption of $r = {\tilde r}$ holds only within a particular
parametrization of the $b\to d$ penguin, so that the CKM ambiguity is
lifted.

There are, in fact, other methods where an assumption can be used to
measure the weak phase of the $t$-quark contribution to the $b\to d$
penguin. My collaborators and I are currently examining such methods.

To summarize: if the unitarity triangle as constructed from
measurements of the CP angles $\alpha$, $\beta$ and $\gamma$ disagrees
with that constructed from measurements of the sides, we may deduce
that there is new physics in $\bd$-$\bdbar$ mixing. However, it may be
that the discrepancy is due not to the presence of new physics, but
rather to an underestimate of the theoretical uncertainties which
enter into the constraints on the unitarity triangle. For this reason,
it is preferable to have direct tests for new physics. 

There are, in fact, several such direct tests, but they all probe new
physics in the $b\to s$ FCNC. One possibility of searching for new
physics in the $b\to d$ FCNC is the following: in the SM the weak
phase of $\bd$-$\bdbar$ mixing is $-2\beta$, while that of the
$t$-quark contribution to the $b\to d$ penguin is $-\beta$. A
comparison of these two weak phases might reveal new physics in the
$b\to d$ FCNC.

Unfortunately, due to the ambiguity in parametrizing the $b\to d$
penguin, it is impossible to cleanly measure the weak phase of the
$t$-quark contribution to the $b\to d$ penguin. In order to measure
this phase, it is necessary to make an assumption about the hadronic
parameters. I presented one example involving the two decays $\bd(t)
\to K^0\kbar$ and $\bs(t) \to\phi\ks$, but there are other methods.
With such an assumption it is possible to detect the presence of new
physics in the $b\to d$ FCNC.

\bigskip
\centerline{\bf Acknowledgments}
\bigskip
I would like to thank the organizers of MRST '99 for a very enjoyable
conference. This research was financially supported by NSERC of Canada and
FCAR du Qu\'ebec.


\begin{references}
  
\bibitem{BCPasym} For a review of CP violation in the $B$ system, see,
  for example, {\it The BaBar Physics Book}, eds.\ P.F. Harrison and
  H.R. Quinn, SLAC Report 504, October 1998.

\bibitem{Wolfenstein} L. Wolfenstein, \prl{51}{83}{1945}.
  
\bibitem{PDG} C. Caso et al.\ (Particle Data Group), \epjc{3}{98}{1}.

\bibitem{AliLon} A. Ali and D. London, hep-ph/9903535, to be published
  in the {\it Eur.\ Phys.\ J.} {\bf C}, 1999.
  
\bibitem{NPpenguins} Y. Grossman and M.P. Worah, \plb{395}{97}{241};
  D. London and A. Soni, \plb{407}{97}{61}.
  
\bibitem{NPBmixing} C.O. Dib, D. London and Y. Nir,
  \ijmp{6}{91}{1253}.

\bibitem{Bnewphysics} For a review of new-physics effects in CP
  asymmetries in the $B$ system, see M. Gronau and D. London,
  \prd{55}{97}{2845}, and references therein.
  
\bibitem{Dalitz} A.E. Snyder and H.R. Quinn, \prd{48}{93}{2139}.

\bibitem{BtoDK} M. Gronau and D. Wyler, \plb{265}{91}{172}. See also
  M.  Gronau and D. London, \plb{253}{91}{483}; I. Dunietz,
  \plb{270}{91}{75}.  Improvements to this method have recently been
  discussed by D. Atwood, I.  Dunietz and A. Soni, \prl{78}{97}{3257}.
  
\bibitem{NirSilv} Y. Nir and D. Silverman, \npb{345}{90}{301}.
  
\bibitem{BstoDsK} R. Aleksan, I. Dunietz, B. Kayser and F. Le
  Diberder, \npb{361}{91}{141}; R. Aleksan, I. Dunietz and B. Kayser,
  \zpc{54}{92}{653}.

\bibitem{penguins} D. London and R. Peccei, \plb{223}{89}{257}.

\bibitem{ucquark} A.J. Buras and R. Fleischer, \plb{341}{95}{379}.
  
\bibitem{LSS} D. London, N. Sinha and R. Sinha, hep-ph/9905404, to be
  published in {\it Phys.\ Rev.} {\bf D}, 1999.
  
\bibitem{isospin} M. Gronau and D. London, \prl{65}{90}{3381}. 
  
\bibitem{helicity} I. Dunietz, H.R. Quinn, A. Snyder, W. Toki and H.J.
  Lipkin, \prd{43}{91}{2193}.
  
\bibitem{KLY2} C.S. Kim, D. London and T. Yoshikawa, hep-ph/9904311,
  to be published in {\it Phys.\ Lett.} {\bf B}, 1999.

\end{references}
\end{document}